\documentclass[pdflatex,sn-basic]{sn-jnl}

\usepackage{graphicx}%
\usepackage{multirow}%
\usepackage{amsmath,amssymb,amsfonts}%
\usepackage{amsthm}%
\usepackage[title]{appendix}%
\usepackage{xcolor}%
\usepackage{booktabs}%
\usepackage{bm}
\usepackage{algorithm}%
\usepackage{algorithmicx}%
\usepackage{algpseudocode}%
\usepackage{listings}%
\usepackage{dsfont}
\usepackage{subcaption}
\let\orcidlogo\relax
\usepackage{orcidlink}

\theoremstyle{thmstyleone}%
\theoremstyle{thmstyletwo}%
\newtheorem{remark}{Remark}%

\theoremstyle{thmstylethree}%
\newtheorem{definition}{Definition}%

\begin{document}

\title[Testing Diagonal Anisotropy]{Testing Diagonal Anisotropy Based on Single Realisation of Spherical Random Field}


\author[1]{\fnm{Shahid} \sur{Khan}\orcidlink{0009-0008-5402-1938}}\email{shahid.khan@latrobe.edu.au}
\equalcont{These authors contributed equally to this work.}

\author*[1]{\fnm{Andriy} \sur{Olenko} \orcidlink{0000-0002-0917-7000}}\email{a.olenko@latrobe.edu.au}
\equalcont{These authors contributed equally to this work.}

\affil*[1]{\orgdiv{Department of Mathematical and Physical Sciences}, \orgname{La Trobe University}, \orgaddress{ \city{Melbourne}, \postcode{3086}, \state{VIC}, \country{Australia}}}


\abstract{Spherical data appear in various applications, including cosmology and the Earth sciences. A standard statistical model for such data employs isotropic spherical random fields. The isotropy assumption may be unrealistic for many real-world datasets. Also, in various applications such as cosmology, only a single realisation of the field is observed, making isotropy impossible to assess through repeated sampling.
This paper considers an alternative anisotropic model, referred to as diagonal anisotropy. To test isotropy against diagonal anisotropy, we develop score and cumulative-sum-type tests. Their asymptotic properties and Monte Carlo-based alternatives suitable for moderate sample sizes are established. The performance of the proposed methods is illustrated by numerical studies via applications to simulated data and to actual Planck cosmic microwave background radiation observations. The tests are also applied to three Planck SMICA cosmic microwave background radiation maps and demonstrate how results can vary for different data releases and map construction procedures.}





\maketitle

\section{Introduction}\label{p3sec1}

Spherical data appear in numerous applications, including cosmology, geosciences, directional and shape data analysis and related fields, see, for example,~\cite{christakos2012, malyarenko2012, duque2024, christophe2017}. One of the classical frameworks for modelling such data is spherical random fields whose index set is the surface of a sphere~(\cite{marinucci2011, yadrenko1983}).

The majority of classical models and statistical methods for spherical random fields rely on the assumption of isotropy, namely, the invariance of their second-order statistical properties under rotations of the sphere. Under isotropy, the spherical harmonic coefficients of a field are uncorrelated, and their variances depend only on the multipole index. The statistical properties of spherical harmonic coefficients and efficient simulation methods for isotropic fields have been studied extensively. See, for example, \cite{marinucci2011, yadrenko1983, lang2015}.

Despite their simplicity and mathematical convenience, isotropic models impose a strong structural assumption whose validity is often questionable in applications. This has motivated the development of several specific anisotropic models. See, for example, \cite{Stein2007, Jun2008, Porcu2018}.

One of the main challenges in the analysis of spherical random fields arises when the field is observed only once. Thus, each spherical harmonic coefficient can be computed only once. Also, statistical methods for spherical data differ substantially from time series. Because they are observed over a fixed, compact domain, increasing spatial resolution adds information through higher multipole degrees rather than an expanding observation domain.

Under isotropy, this lack of statistical replication can be partially overcome. All harmonic coefficients associated with a given multipole degree have the same variance, so the collection of coefficients at that degree acts as a set of replicates for estimating the angular power spectrum. However, for a general anisotropic field, the covariance structure contains too many unknown quantities to be estimated reliably from a single realisation. Therefore, additional structural assumptions are essential for identifiability and valid statistical inference.

This issue is particularly important in the cosmic microwave background (CMB) analysis, where only one full-sky map is observed~(see Appendix~\ref{secA1}). Figure~\ref{p3image1} shows an example of a CMB map from the Planck space mission. Several large-scale features, including alignments among low-degree multipoles~(\cite{de2004}), the so-called “axis of evil”~(\cite{land2005}), and hemispherical asymmetry in the fluctuation amplitude~(\cite{eriksen2004}), suggest possible departures from isotropy with a preferred directional structure, rather than arbitrary anisotropy. Its consistent appearance in both WMAP and Planck space mission data has made an instrumental explanation less plausible~(\cite{copi2015, planck2020}). These types of anomalies have been investigated using several approaches, including low-degree multipole maps~(\cite{bielewicz2004}), geometric methods~(\cite{Weeks2004, Leonenko2021, BROADBRIDGE_2022}), multipole vectors~(\cite{copi2004}), refined estimators of the axis-of-evil statistic~(\cite{land2007}), and correlations with the ecliptic plane and the CMB dipole~(\cite{Schwarz2004}).

\begin{figure}[htbp]
\centering
\includegraphics[width=0.8\textwidth]{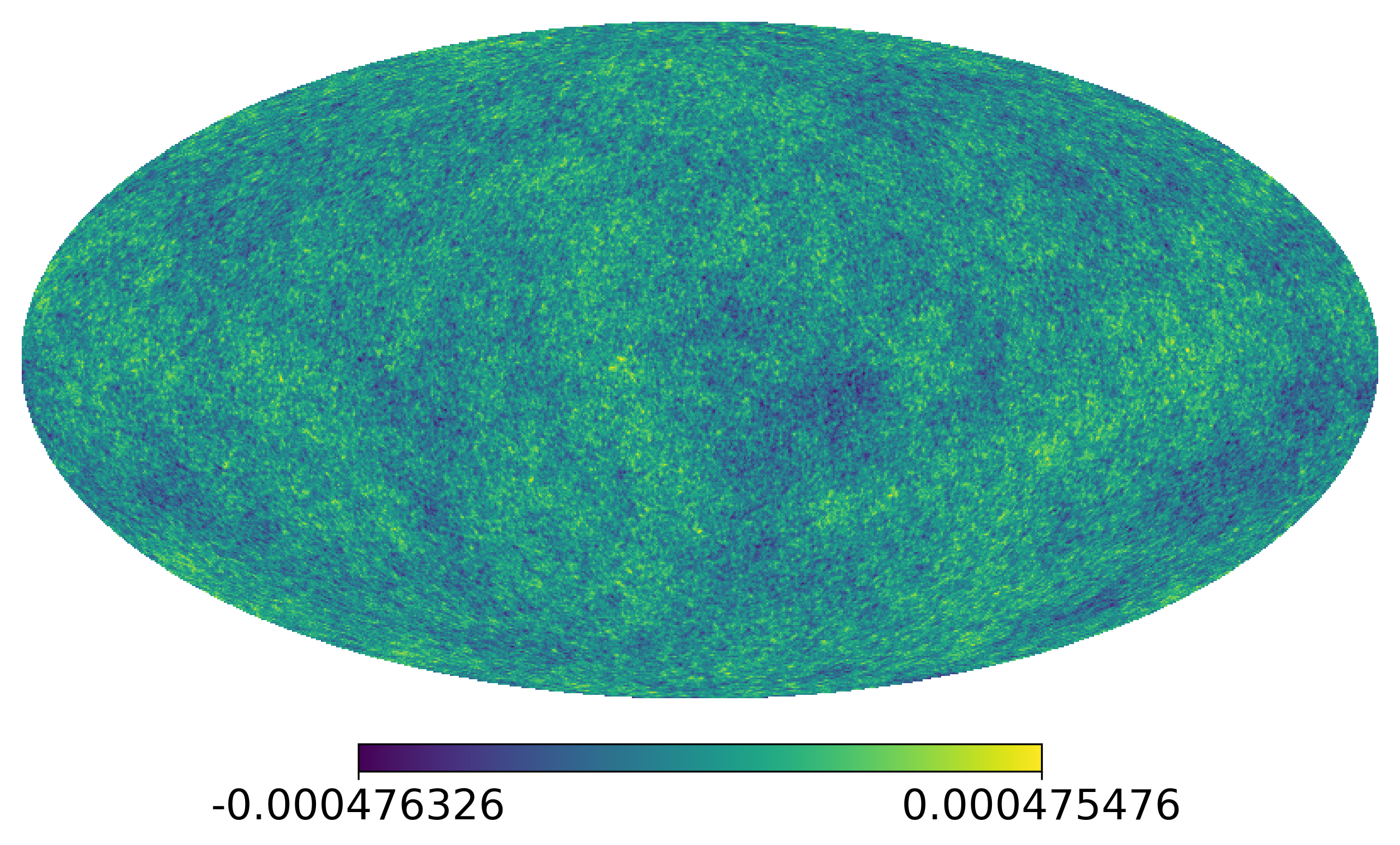}
\caption{Full-sky Planck SMICA CMB temperature map in Mollweide
projection.}\label{p3image1}
\end{figure}

A classification of anisotropic scenarios and initial framework for a general theory of anisotropic random fields on the sphere was developed in~\cite{Durastanti2026}. In particular, it considered diagonal anisotropy, where harmonic coefficients are uncorrelated for distinct multipole degrees and azimuthal orders, while their variances may depend on both indices. Diagonality alone, however, does not resolve the inferential difficulty, as each variance can be estimated only from one coefficient. Therefore, a semiparametric model is introduced, in which the variances share a common modulation function. Under this formulation, testing isotropy reduces to testing a specific form of homoscedasticity with the common constant function at different multipoles.

This paper develops two methods for testing such hypotheses, derives related asymptotic distributions and studies statistical properties of the tests as the number of used multipole degrees increases. Also, Monte Carlo procedures are proposed for scenarios with a moderate number of degrees. Numerical studies based on simulated and actual CMB data illustrate the developed methods.


\section{Theoretical background}\label{p3sec2}

This section summarises some key facts about isotropic and anisotropic random fields, that are required for the discussion that follows. More details can be found in \cite{marinucci2011, Durastanti2026}.

Let $T(\cdot) $ be a second-order zero-mean Gaussian random field on the unit sphere
$\mathbb S^2:= \{\mathbf{x} \in \mathbb{R}^3,\, ||\mathbf{x}||=1 \}$, with the spherical harmonic expansion that holds in $L^2(\Omega\times\mathbb S^2)$
\[
T (\mathbf{x})
=
\sum_{\ell=0}^{\infty}\sum_{m=-\ell}^{\ell}
a_{\ell m}Y_{\ell m}(\mathbf{x}),
\]
where $\{ {Y}_{\ell m}(\mathbf{x})\}$ is an orthonormal complex spherical-harmonic basis of $L^2(\mathbb S^2)$ and $\{ a_{\ell m}\}$ are the associated random harmonic coefficients.  Under the Gaussian-field assumption, the collection $\{a_{\ell m}\}$ is jointly Gaussian.

If a realisation of $T(\cdot)$ is observed over the sphere, its harmonic
coefficients can, in principle, be recovered as
\[
a_{\ell m}
=
\int_{\mathbb S^2}
T (\mathbf{x})\overline{Y_{\ell m}(\mathbf{x})}\,d\mathbf{x}.
\]

Therefore, a continuous full-sky observation over the entire sphere $\mathbb S^2$ at a fixed time can yield only one value of each coefficient $a_{\ell m}$. It does not give repeated realisations of the coefficients
$a_{\ell m}$.

Let $SO(3)$ denote the group of rotations on $\mathbb{R}^3.$
\begin{definition}\label{isotropy}
The zero-mean random field $T(\mathbf{x})$ is called isotropic if, for all $\mathbf{x}_{1}, \mathbf{x}_{2} \in \mathbb{S}^2$ and all rotations $\rho \in SO(3),$ it holds \[ C(\mathbf{x}_1,\mathbf{x}_2):=\mathbb {E}(T(\mathbf{x}_1) \overline{T(\mathbf{x}_2)})=C(\rho \mathbf{x}_1,\rho \mathbf{x}_2).\]
\end{definition}

For a centred second-order isotropic field,
\[
\mathbb E\!\left(
a_{\ell m}\overline{a_{\ell' m'}}
\right)
=
C_\ell\delta_{\ell\ell'}\delta_{mm'},
\]
where $\delta _{\ell\ell^{\prime }}$ is the Kronecker delta function and $C_\ell$ is the angular power spectrum at the multipole $\ell\in \mathbb{N}_0:=\mathbb{N} \cup \{0\}.$  The values $C_\ell$ are the same for
every $m=-\ell,\ldots,\ell$. Thus, one can use the following unbiased estimator
\[
\widehat C_\ell
:=
\frac{1}{2\ell+1}
\sum_{m=-\ell}^{\ell}|a_{\ell m}|^2.
\]

For a real-valued isotropic Gaussian field,
$
\operatorname{Var}(\widehat C_\ell)
=
2({2\ell+1})^{-1}{C_\ell^2}.
$
This irreducible single-realisation uncertainty is usually called cosmic
variance.

\begin{definition}
For a general random field on $\mathbb{S}^2$ with spherical harmonic coefficients $\{a_{\ell m}\}$, the  angular power spectrum is defined as
\begin{equation*} \label{Cllmm} C_{\ell \ell^{\prime} m m^{\prime}}:=\mathbb {E}\left( a_{\ell m} \overline{a_{\ell^{\prime} m^{\prime}}}\right),\quad \ell, \ell^{\prime} \in \mathbb{N}_0, \ m \in \{-\ell, \ldots, \ell\},\ m^{\prime} \in \{-\ell^{\prime}, \ldots, \ell^{\prime}\}.
\end{equation*}
\end{definition}

Thus, anisotropy occurs if and only if there exist indices $\ell,\ell^{\prime},m,m^{\prime}$ such that either
\begin{equation*} \label{Ee}
    C_{\ell \ell^{\prime} m m^{\prime}}  \neq 0, \quad \mbox{when} \quad |\ell - \ell^{\prime}| + |m - m^{\prime}| \neq 0,
\end{equation*}
or
\begin{equation*}
    C_{\ell \ell m m} \neq C_{\ell \ell m^{\prime} m^{\prime}},
    \quad \text{for some } m\neq m^{\prime}.
\end{equation*}

The angular spectrum satisfies Hermitian symmetry
\[
C_{\ell \ell^{\prime} m m^{\prime}} = \overline{C_{\ell^{\prime} \ell m^{\prime} m}},
\quad \ell, \ell^{\prime} \in \mathbb{N}_0, \quad
m \in \{-\ell, \ldots, \ell\}, \quad m^{\prime} \in \{-\ell^{\prime}, \ldots, \ell^{\prime}\},
\]
which follows from the definition of covariance for complex-valued harmonic coefficients.

For a real-valued field,
$
a_{\ell (-m)}=(-1)^m\overline{a_{\ell m}},
$
and hence
$
|a_{\ell (-m)}|^2=|a_{\ell m}|^2.
$
Thus, for the same $\ell,$ the coefficients at $m$ and $-m$ should not be treated as independent replicates.
If the field is real-valued,
\[
C_{\ell \ell^{\prime} m m^{\prime}} = (-1)^{m+m^{\prime}} \overline{C_{\ell\ell^{\prime} (-m) (-m^{\prime})}}.
\]

\begin{definition} \label{def22}
    A random field is diagonally anisotropic if
    \[
    C_{\ell \ell^\prime m m^\prime} = \delta_{\ell\ell^\prime} \delta_{m m^\prime}\, {C}_{\ell m}
    \]
    with
    \begin{equation*} \label{dan}
        {C}_{\ell m} \neq {C}_{\ell m^{\prime}} \quad \text{for some } \ell \in \mathbb{N}_0, \ m \neq m^{\prime}, \ m, m^\prime \in \{-\ell, \ldots, \ell\}.
    \end{equation*}
\end{definition}

For real-valued fields,
$
C_{\ell(-m)}=C_{\ell m}.
$

\begin{definition}\label{def_axial}
    A random field is axially symmetric around the axis defined by the preferred direction $\widehat{\bm n},$ which passes through the point~$\mathbf{x}_0$ and its antipode if, for all $\mathbf{x}_1, \mathbf{x}_2 \in \mathbb{S}^2$ and $\alpha \in [0, 2\pi)$, its covariance function satisfies the relation
    \[ C(\rho_{\mathbf{x}_0, \alpha} \mathbf{x}_1, \rho_{\mathbf{x}_0, \alpha} \mathbf{x}_2) = C(\mathbf{x}_1, \mathbf{x}_2),
    \]
    where $\rho_{\mathbf{x}_0, \alpha}$ denotes a rotation around that axis through $\mathbf{x}_0$ by the angle $\alpha$.
\end{definition}

\begin{remark}
More details about these fields and related references can be found in the paper~\cite{Durastanti2026}, which investigated the spectral and related properties of these fields and established their connection with convolution fields generated from isotropic fields.
\end{remark}

If the covariance is invariant under rotations about the $z$-axis,
the covariance of harmonic coefficients has the form
$
\mathbb E\!\left(
a_{\ell m}\overline{a_{\ell'm'}}
\right)
=
\delta_{mm'}\,C^m_{\ell\ell'}.
$
Thus, axial symmetry implies diagonality in $m$, but not generally in
$\ell$, see \cite{Emery2019}.
Therefore, the diagonally anisotropic models
form a subclass of axially symmetric models. In this subclass,  the covariance of the harmonic coefficients is additionally assumed to be diagonal in $\ell$.

\section{Structured non-isotropic model}\label{p3sec3}
For an arbitrary random field with the angular power spectrum $\{C_{\ell m}\}$, the immediate estimator
$
\widehat C_{\ell m}=|a_{\ell m}|^2
$
is unbiased, but it contains only one observation for each unknown variance.
Without additional
assumptions, the coefficient $C_{\ell m}$ cannot generally be estimated
consistently from one realisation.
Estimation becomes possible when several multipoles contain information about a
lower-dimensional object.

The possible preferred direction motivates models in which the variance of the harmonic coefficients varies for azimuthal orders relative to a fixed axis. Broadly, two approaches have been proposed in the literature. The first specifies a parametric alternative and statistical tests to detect it. For example, the bipolar spherical harmonic formalism decomposes the covariance structure into isotropic and anisotropic components. It tests whether the anisotropic components vanish~(\cite{Souradeep2006}). Some of the available methods include quadratic maximum-likelihood and minimum-variance estimators, and Bayesian approaches for direction-dependent power spectra~(\cite{pullen2007, hanson2009, groeneboom2009, kim2013}). However, such tests have low power for misspecified alternatives.

The second approach imposes a weaker symmetry. Axial symmetry, for example, requires invariance only under rotations about a fixed axis. Under this assumption, the covariance matrix of the harmonic coefficients is diagonal with respect to azimuthal order, while dependence between different multipole degrees is still permitted~(\cite{jones1963, Stein2007, Jun2008}). Semiparametric classes of axially symmetric covariance functions on the sphere have been studied in~\cite{Emery2019}. Tests for isotropy and axial symmetry based on subsampling and periodogram methods were proposed in~\cite{Guan2004, scaccia2005}. However, these procedures often rely on large-area asymptotics or on spatial replication.

This paper considers the following semiparametric model
with a direct angular-momentum normalisation and its alternative parametrisation in terms of $m/\ell:$
\[
C_{\ell m}
=
C_\ell
\tilde g\!\left(
\frac{m}{\sqrt{\ell(\ell+1)}}
\right)=
C_\ell
g\!\left(
\frac{m}{\ell}
\right), \qquad \ell\geq1,\quad -\ell\leq m\leq\ell.
\]

It is assumed that random fields are non-degenerate and real-valued, with unknown scale parameters $C_\ell>0$ and an unknown function $g:[-1,1]\to(0,\infty)$. For a real-valued field, the function $g(\cdot)$ should be even.
 It is convenient to use the~$m/\ell$ scale instead of  ${m}/\sqrt{\ell(\ell+1)},$ because it gives a common domain $[-1,1]$.
If $\ell$ is sufficiently large $\sqrt{\ell(\ell+1)}\approx \ell.$ Also, regardless of the type of these functions, the proposed methods will test the hypothesis of a constant function $\tilde g(\cdot)$ or $g(\cdot).$

 To obtain an $L^2(\Omega\times\mathbb S^2)$ field, one needs to assume
\[
\sum_{\ell=1}^{\infty}C_\ell\sum_{m=-\ell}^{\ell}g(m/\ell)<\infty.
\]
Since the positive-valued function $g(\cdot)$ is bounded above and away from zero on $[-1,1]$, this is equivalent to the standard assumption on the angular power spectrum~$\sum_{\ell\geq1}(2\ell+1)C_\ell<\infty$.

In the considered model,
the decomposition is not uniquely identified. One can impose the identifiability normalisation condition
$\int_0^1g(x)\,dx=1.
$
 Then, the coefficients~$\{C_\ell\}$ control average power, while the function $g(\cdot)$ controls
its directional redistribution. Note that this continuous normalisation is convenient but does not make $C_\ell$ equal to the exact $\ell$-arithmetic average of $C_{\ell m}$. The latter involves the discrete weights induced by the harmonic multiplicities. Notice that the considered later hypotheses and tests do not require this normalisation, because $C_\ell$ will be  profiled out for every~$\ell$.

For the CMB applications, the model has the following motivation. Choose the $z$-axis to coincide with a supposed
preferred direction $\widehat{\bm n}$. The model
\[
\mathbb E\!\left(
a_{\ell m}\overline{a_{\ell'm'}}
\right)
=
C_\ell
\tilde g\!\left(
\frac{m}{\sqrt{\ell(\ell+1)}}
\right)
\delta_{\ell\ell'}\delta_{mm'}
\]
describes an axially aligned redistribution of spectral power among the $m$-modes, while retaining independence for different $\ell$. The $\ell=0$ mode carries no information about $g(\cdot)$ and is excluded from the testing procedure.

The angular-momentum interpretation provides a geometric motivation. For a
spherical harmonic,
$
|\bm L|^2=\ell(\ell+1)$ and the projection is $
L_z=m,
$
so the standard angular-momentum eigenvalue in {\cite{Varshalovich1988}} give
\[
\frac{m}{\sqrt{\ell(\ell+1)}}
=
\frac{L_z}{|\bm L|}.
\]
Heuristically, at high frequency, this quantity can be interpreted as an orientation
coordinate relative to the preferred axis. Hence $\tilde{g}(\cdot)$ describes
orientation-dependent spectral power.

\section{Testing isotropy against diagonal anisotropy}\label{p3sec4}
This section introduces and studies two test procedures for isotropy against the specified diagonal axially aligned alternatives.

The preferred axis (and hence the coordinate system defining $m$) is assumed to be selected in advance,  independently of the tested coefficients. In this and following sections, it is assumed that the field is real-valued and Gaussian and the covariance of the harmonic coefficients is diagonal in $(\ell,m)$. Except the conjugacy relation, $a_{\ell (-m)}=(-1)^m\overline{a_{\ell m}}$, the other coefficients indexed by $\ell\geq 1$ and $0\leq m\leq\ell$ are independent for distinct index pairs.

Consider the hypotheses
\[
H_0: g(x)\equiv const
\quad\text{against}\quad
H_1: g(x)\not\equiv const.
\]
Under the mentioned normalisation condition, $H_0$ may also be written as $g(x)\equiv 1$.

 Under $H_0$,
$
a_{\ell m}\sim\mathcal{N}(0,const \cdot C_\ell)
$.
If $m=0,$ the harmonic coefficient is real, whereas for $m>0$ the coefficients are proper complex Gaussian random variables.  Therefore, \[|a_{\ell0}|^2/(const \cdot C_\ell)\sim\chi_1^2\quad \mbox{and}\quad 2|a_{\ell m}|^2/(const \cdot C_\ell)\sim\chi_2^2, \quad \mbox{for}\quad m\not = 0.\]

In the next derivations we use the real-field assumption. For each $\ell\geq 1$, consider  only $m=0,1,\ldots,\ell,$ and define
\[
d_m :=
\begin{cases}
    1, & m=0,\\
    2, & m>0.
\end{cases}
\]

For each $(\ell, m)$ define
\begin{equation}\label{Plm}
         P_{\ell m}
        :=
        \frac{d_m |a_{\ell m}|^2}
        {\displaystyle\sum_{i=0}^{\ell}d_i|a_{\ell i}|^2}
        =
        \frac{d_m |a_{\ell m}|^2}
        {(2\ell+1)\widehat C_\ell},
\end{equation}
so the nuisance scale $C_\ell$ cancels in $P_{\ell m}.$

Note that by Gaussianity,
$(P_{\ell 0},\ldots,P_{\ell \ell})
\sim
\operatorname{Dirichlet}
\left(1/2,1,\ldots,1\right)
$ and it holds $
\sum_{m=0}^{\ell}P_{\ell m}=1
$.  Refer to~\cite{kai2011} for a detailed discussion of Dirichlet distributions and their properties used later.

Under $H_0$, there should be no systematic relationship between
$P_{\ell m}$ and $m/\ell$. This relationship can therefore be assessed using appropriate tests of homoscedasticity.

\subsection{Score test}\label{p3subsec4a}

{\bf Theoretical derivation of the test}. To test against a general nonconstant variance function, one can consider the model
\[
\log g_{\bm\theta}(x)
=
\theta_1h^{(1)}(x)+
\theta_2h^{(2)}(x)+
\ldots+
\theta_Kh^{(K)}(x)=\bm\theta^{\mathsf T}\bm h(x),
\]
where $\bm\theta:=(\theta_1,
\ldots,
\theta_K)^{\mathsf T}$ and $\bm h(x):=\bigl(h^{(1)}(x),\ldots,h^{(K)}(x)\bigr)^{\mathsf T}.$ The basis functions  $h^{(k)}(x),$ $x\in[0,1],$  $k=1,...,K,$ and their number $K\ge 1,$ are fixed in advance.
For example, one may use a polynomial power basis. The orthogonal shifted Legendre polynomials on $[0,1]$ are usually
preferable numerically.

We will use the notations
\[
\overline{\bm h}:=\int_0^1\bm h(x)\,dx \qquad\mbox{and}\qquad
\bm\Sigma_h
:=
\int_0^1
\left(\bm h(x)-\overline{\bm h}\right)
\left(\bm h(x)-\overline{\bm h}\right)^{\mathsf T}\,dx.
\]
We assume that the matrix $\bm\Sigma_h$ is positive definite. The last condition excludes degenerate cases, for example, linear combinations of the selected basis functions.

For each basis function, define
\begin{equation}\label{tildeh}
\widetilde h^{(k)}_{\ell m}
:=
h^{(k)}(m/\ell)
-
\frac1{2\ell+1}
\sum_{i=0}^\ell d_i h^{(k)}(i/\ell).
\end{equation}

In this setting the hypothesis $H_0$ becomes $\theta_1=
\theta_2=
\ldots=
\theta_K=0.$

Define $n_\ell:=2\ell+1$ and
$R_{\ell m}:=n_\ell P_{\ell m}/d_m$. Let us use the finite number $L$ of multipoles $\ell=1,...,L.$

Removing terms that do not depend on the parameters $(C_\ell,\bm\theta)$, the log-likelihood contribution of the multipole $\ell$ is
\[\mathcal L_\ell(C_\ell,\bm\theta)
=-\frac12\sum_{m=0}^{\ell}
\left(d_m\log\left(C_\ell g_{\bm\theta}(m/\ell)\right)
+\frac{d_m|a_{\ell m}|^2}{C_\ell g_{\bm\theta}(m/\ell)}\right).
\]
For fixed $\bm\theta$, its scale parameter maximiser is
\[
\widehat C_\ell(\bm\theta)
:=\frac1{n_\ell}\sum_{m=0}^{\ell}
\frac{d_m|a_{\ell m}|^2}{g_{\bm\theta}({m/\ell})}.
\]
Differentiating the log-likelihood contribution from multipole $\ell$ at $\bm\theta=\bm 0$ gives
\[
\left.\frac{\partial\mathcal L_\ell(\widehat C_\ell(\bm\theta),\bm\theta)}{\partial\theta_k}\right|_{\bm\theta=\bm0}
=\frac12\sum_{m=0}^{\ell}d_m(R_{\ell m}-1)h^{(k)}(m/\ell).
\]
Because $\sum_m d_m(R_{\ell m}-1)=0$, one can replace $h^{(k)}(\cdot)$ with its weighted-centred version $\widetilde h^{(k)}_{\ell m}(\cdot)$, keeping the score unchanged.  Using the independence of multipoles, by~\eqref{tildeh}, the components of the score vector are
\[
U_k
:=
\frac12
\sum_{\ell=1}^{L}\sum_{m=0}^{\ell}
d_m\bigl(R_{\ell m}-1\bigr)\widetilde h^{(k)}_{\ell m}
=
\frac12
\sum_{\ell=1}^{L}(2\ell+1)
\sum_{m=0}^{\ell}
P_{\ell m}\widetilde h^{(k)}_{\ell m}.
\]

Under $H_0$, \eqref{tildeh} gives
$
\mathbb E_0(U_k)=0.
$

For a Dirichlet vector,
\[
\operatorname{Cov}_0(P_{\ell m},P_{\ell j})
=\frac{d_m\left(n_\ell\mathbf \delta_{mj}-d_j\right)}
{n_\ell^2(n_\ell/2+1)}.
\]
The centring in \eqref{tildeh} gives $\sum_{m=0}^{\ell} d_m\widetilde h^{(k)}_{\ell m}=0$. Hence, the covariance
\begin{align*}
\operatorname{Cov}_0 & \left((n_\ell/2)\sum_{m=0}^{\ell} P_{\ell m}\widetilde h^{(k_1)}_{\ell m}, (n_\ell/2)\sum_{m=0}^{\ell} P_{\ell m}\widetilde h^{(k_2)}_{\ell m}\right)\\
&=\frac{1}{2(n_\ell+2)}
\sum_{m=0}^{\ell}\sum_{m'=0}^{\ell}\left(d_m n_\ell\mathbf \delta_{mm'}-d_md_{m'}\right) \widetilde h^{(k_1)}_{\ell m}
\widetilde h^{(k_2)}_{\ell m'}\\
&=\frac{n_\ell}{2(n_\ell+2)}
\sum_{m=0}^{\ell}d_m\widetilde h^{(k_1)}_{\ell m}
\widetilde h^{(k_2)}_{\ell m}.
\end{align*}

Because different multipoles are independent under the hypothesis~$ H_0$, the covariance matrix of the score vector is
\[V_{k_1k_2}
:=\operatorname{Cov}_0(U_{k_1},U_{k_2})
=
\sum_{\ell=1}^{L}
\frac{2\ell+1}{2(2\ell+3)}
\sum_{m=0}^{\ell}
d_m\widetilde h^{(k_1)}_{\ell m}
\widetilde h^{(k_2)}_{\ell m}.
\]

Let
$
\bm U:=(U_1,\ldots,U_K)^{\mathsf T},
$ $
\bm V:=(V_{k_1k_2})_{k_1,k_2=1}^K.
$ Assume that $\bm V$ is nonsingular for the selected $L.$ This holds for all sufficiently large $L$ under the stated assumptions.
Then, one can use the corresponding statistic
$
    Q
    =
    \bm U^{\mathsf T}
    \bm V^{-1}
    \bm U.
$

Let $\bm U^{(\ell)}:=(U_1^{(\ell)},\ldots,U_K^{(\ell)})^{\mathsf T}$ be the score contribution from multipole $\ell$. Therefore, $\bm U=\sum_{\ell=1}^L\bm U^{(\ell)}$. The vectors $\bm U^{(\ell)}$ are independent and centred. For every fixed $\bm c\in\mathbb R^K$, the Dirichlet moment formula gives,
\[
\mathbb E_0|\bm c^{\mathsf T}\bm U^{(\ell)}|^4=O(n_\ell^2).
\]
Also,
$\bm V/(L(L+2))\to\bm\Sigma_h/2$. Therefore,
$\bm c^{\mathsf T}\bm V\bm c\asymp L^2$ for $\bm c\ne\bm 0$. Hence, the Lyapunov ratio is  bounded as
\[
\frac{\sum_{\ell=1}^L\mathbb E_0|\bm c^{\mathsf T}\bm U^{(\ell)}|^4}
{(\bm c^{\mathsf T}\bm V\bm c)^2}
=O\!\left(\frac{\sum_{\ell=1}^L n_\ell^2}{L^4}\right)
=O(L^{-1})\to 0,
\]
when $L\to \infty.$

Therefore, one can apply Lyapunov's central limit theorem~\cite{billingsley1995probability}, which yields
$\bm V^{-1/2}\bm U\Rightarrow\mathcal N(\bm 0,\bm I_K)$. Finally, the continuous mapping theorem \cite{Billingsley1999} gives
that under $H_0$, when the number of independent multipoles
increases, asymptotically,
$
    Q\sim\chi_K^2.
$

Hence, the asymptotic $p$-value of the score test is
\[
p_s
:=
\mathbb P\!\left(\chi_K^2\ge Q_{\mathrm{obs}}\right),
\]
where $ Q_{\mathrm{obs}}$ is computed using $P_{\ell m}^{\mathrm{obs}}$ values obtained via \eqref{Plm} using $a_{\ell m}^{\mathrm{obs}}$ for a given realisation of the random field $T(\cdot).$

\medskip

\noindent{\bf Consistency of the score test.} Consider the large-$L$ behaviour of the score statistic.

The alternative of $H_0$ can be written in the form
$
g_{\bm\theta}(x)
=
\exp\left(\bm\theta^{\mathsf T}\bm h(x)\right),
$ $\bm\theta \not =\bm 0.$
 Define
\[
\bm\mu(\bm\theta)
:=
\frac{\int_0^1 \bm h(x)
    \exp\left(\bm\theta^{\mathsf T}\bm h(x)\right)\,dx}
{\int_0^1
    \exp\left(\bm\theta^{\mathsf T}\bm h(x)\right)\,dx},
\qquad
\bm\Delta(\bm\theta)
:=
\bm\mu(\bm\theta)-\bm\mu(\bm0).
\]

Note that for $m=0,\ldots,\ell$, one can write
\[
d_m|a_{\ell m}|^2
=
C_\ell g_{\bm\theta}(m/\ell)Z_{\ell m},
\]
where $Z_{\ell m}\sim\chi^2_{d_m}$ and independent.

Then,
\[
\sum_{m=0}^{\ell}P_{\ell m}\bm h(m/\ell)
=\frac{n_\ell^{-1}\sum_{m=0}^{\ell}g_{\bm\theta}(m/\ell)Z_{\ell m}\bm h(m/\ell)}
{n_\ell^{-1}\sum_{m=0}^{\ell}g_{\bm\theta}(m/\ell)Z_{\ell m}}.
\]
The variances of both the numerator and denominator are $O(n_\ell^{-1})$, because $g_{\bm\theta}(\cdot)$ and~$\bm h(\cdot)$ are bounded and $\operatorname{Var}(Z_{\ell m})=2d_m$. Thus, the weighted Riemann sums converge
\begin{align*}
n_\ell^{-1}\sum_{m=0}^{\ell}g_{\bm\theta}(m/\ell)Z_{\ell m}\bm h(m/\ell)
&\overset p\longrightarrow \int_0^1g_{\bm\theta}(x)\bm h(x)\,dx,\\
n_\ell^{-1}\sum_{m=0}^{\ell}g_{\bm\theta}(m/\ell)Z_{\ell m}
&\overset p\longrightarrow \int_0^1g_{\bm\theta}(x)\,dx, \quad \mbox{when}\ \ell\to\infty,
\end{align*}
as the terms with weight $d_0=1$ are asymptotically negligible.

Let
\[
\overline{\bm h}_\ell
=\frac1{n_\ell}\sum_{m=0}^{\ell}d_m\bm h(m/\ell).
\]
Then,  $\overline{\bm h}_\ell\to\overline{\bm h}$, when $\ell\to\infty.$

Therefore,
\[
\frac{\bm U^{(\ell)}}{n_\ell}
=\frac12\left(\sum_{m=0}^{\ell}P_{\ell m}\bm h(m/\ell)-\overline{\bm h}_\ell\right)
\overset p\longrightarrow\frac12\bm\Delta(\bm\theta).
\]
These vectors are uniformly bounded. Their independence for different $\ell$, together with
$\sum_{\ell=1}^L n_\ell^2/(L(L+2))^2=O(L^{-1})$, shows that the weighted average of their centred versions converges to zero in probability. Toeplitz's lemma applied to their expectations then yields
$\bm U/(L(L+2))\overset p\to\bm\Delta(\bm\theta)/2$.

For the covariance matrix, weighted Riemann-sum convergence gives
\[
\frac1{n_\ell}\sum_{m=0}^{\ell}d_m
\left(\bm h(m/\ell)-\overline{\bm h}_\ell\right)
\left(\bm h(m/\ell)-\overline{\bm h}_\ell\right)^{\mathsf T}
\to\bm\Sigma_h.
\]
Summing the covariances and using
$\max_{1\le \ell\leq L}n_\ell/(L(L+2))\to 0$, one obtains $ {\bm V}/(L(L+2))
\to
\bm\Sigma_h /2,
$
when $L\to\infty.$

Therefore,
\begin{equation}
    \frac{Q}{L(L+2)}
    \overset{p}{\to }
    \frac12
    \bm\Delta(\bm\theta)^{\mathsf T}
    \bm\Sigma_h^{-1}
    \bm\Delta(\bm\theta).
    \label{eq:score-fixed-alt}
\end{equation}

The right-hand side of \eqref{eq:score-fixed-alt} is strictly positive for all $\bm\theta\neq\bm 0$. Indeed, if
\[
\psi(\bm\theta)
:=
\log\left(\int_0^1
\exp\left(\bm\theta^{\mathsf T}\bm h(x)\right)\,dx\right),
\]
then $\bm\mu(\bm\theta)=\nabla\psi(\bm\theta)$ and $\bm\Sigma_h=\nabla^2\psi(\bm0)$. Notice that $\nabla^2\psi(\bm\eta)$ is the covariance matrix of $\bm h(X)$ under the probability density of $X$ proportional to $\exp\left(\bm\eta^{\mathsf T}\bm h(x)\right)$ on $[0,1]$. This density is strictly positive, so positive definiteness at $\bm0$ implies positive definiteness for every $\bm\eta$. Therefore,
\[
\bm\theta^{\mathsf T}
\left(\nabla\psi(\bm\theta)-\nabla\psi(\bm0)\right)
=\int_0^1\bm\theta^{\mathsf T}\nabla^2\psi(t\bm\theta)\bm\theta\,dt>0
\qquad
\text{for }\bm\theta\neq\bm0.
\]
Thus, $\bm\Delta(\bm\theta)\neq\bm0$ and $\bm\Delta(\bm\theta)^{\mathsf T}
    \bm\Sigma_h^{-1}
    \bm\Delta(\bm\theta)>0$.

It follows from (\ref{eq:score-fixed-alt}) that $Q\to\infty$ in probability, when $L\to\infty.$ Therefore, for the considered log-linear model, the score test is consistent against every non-zero alternative.

\medskip

\noindent{\bf Simulated distribution and test.} For small and moderate values of $L$ in numerical studies, a reference null distribution can be generated under $H_0.$
It can be simulated without estimating any $C_\ell$. For each Monte Carlo replication $s=1,...,S,$ and each multipole $\ell=1,\ldots,L$,  draw independently $
(P^{(s)}_{\ell0},\ldots,P^{(s)}_{\ell\ell})
$ from $\operatorname{Dirichlet}(1/2,1\ldots,1)$. Use the same observed set of multipoles and the same $m/\ell $. Recompute $\bm U_s$ and then~$Q_s$. For~$S$ replications, one can compute the corresponding estimated $p$-value of the score test as
\[
    \widehat p_{s,\mathrm{MC}}:=
    \frac{1+\sum_{s=1}^{S}
\mathds{1}\left(Q_s\geq Q_{\mathrm{obs}}\right)}{1+S},
\]
where $\mathds{1}\left(\cdot\right)$ is an indicator function.

Adding 1 corresponds to the observed dataset itself as one of the possible permutations.
This is an exact finite-sample null calibration for the ideal full-sky Gaussian diagonal null model. It remains valid for unequal numbers of available $m$ modes across multipoles because the appropriate Dirichlet dimension changes with $\ell$.

\subsection{CUSUM test}\label{p3subsec4b}
{\bf Theoretical derivation of the test}.
A complementary omnibus test can be obtained by using the same normalised powers $P_{\ell m}$ in \eqref{Plm}.

For $u\in[0,1]$, define
the cumulative normalised power
\begin{equation*}
F_\ell(u):
=
\sum_{m=0}^{\lfloor \ell u\rfloor }P_{\ell m}.
\label{eq:Cell}
\end{equation*}
Under $H_0$, by the properties of the Dirichlet distribution, its exact mean is
\begin{equation*}
p_\ell(u):
=
\mathbb E_0\left(F_\ell(u)\right)
=
\frac{\sum_{m=0}^{\lfloor \ell u\rfloor }d_m}{2\ell+1}
=
\frac{2\lfloor \ell u\rfloor+1}{2\ell+1}.
\label{eq:pell}
\end{equation*}

Consider
\[
A=\sum_{m=0}^{\lfloor\ell \min(u,v)\rfloor}{d_m}/{2},
\qquad
B=\sum_{m=\lfloor\ell \min(u,v)\rfloor+1}^{\lfloor\ell \max(u,v)\rfloor}{d_m}/{2},
\qquad
D={n_\ell}/{2}-(A+B).
\]
The property of Dirichlet aggregation gives
\[
\bigl(F_\ell(\min(u,v)),\,F_\ell(\max(u,v))-F_\ell(\min(u,v)),\,1-F_\ell(\max(u,v))\bigr)
\sim\operatorname{Dirichlet}(A,B,D).
\]

Therefore, under $H_0$, for $u,v\in[0,1],$ the Dirichlet covariance formula yields
\begin{equation}
\operatorname{Cov}_0\left(F_\ell(u),F_\ell(v)\right)
=\frac{A\left(n_\ell/2-(A+B)\right)}{(n_\ell/2)^2(n_\ell/2+1)}=
\frac{2\left(
 p_\ell(\min(u,v))-p_\ell(u)p_\ell(v)
\right)}{n_\ell+2}
.
\label{eq:rowcov-sphere}
\end{equation}

Let us group multipoles using their numbers of actual degrees of freedom, $n_\ell$. Let us define
\begin{equation*}
A_L(u):
=
\sum_{\ell=1}^{L}n_\ell \left(F_\ell(u)-p_\ell(u)\right),
\quad
V_L:
=
\sum_{\ell=1}^{L}
\frac{n_\ell^2}{n_\ell+2}
,
\label{eq:VL}
\end{equation*}
and
\begin{equation}
Z_L(u)
:=
\frac{A_L(u)}{\sqrt{2V_L}}.
\label{eq:ZL}
\end{equation}

The independence of distinct multipoles and \eqref{eq:rowcov-sphere} give
\begin{equation}
\operatorname{Cov}_0\left(Z_L(u),Z_L(v)\right)
=
\frac{1}{V_L}
\sum_{\ell=1}^{L}
\frac{n_\ell^2}{n_\ell+2}
\left(
 p_\ell(\min(u,v))-p_\ell(u)p_\ell(v)
\right).
\label{eq:pooledcov-sphere}
\end{equation}
Note that for $\ell\to \infty$ and $L\to \infty$ it holds
\begin{equation}\label{pu}
\sup_{0\le u\le1}|p_\ell(u)-u|
=O(\ell^{-1})
\quad \mbox{and}\quad
V_L\sim L(L+2).
\end{equation}
Hence, it follows from \eqref{eq:pooledcov-sphere} that
\[
\operatorname{Cov}_0\left(Z_L(u),Z_L(v)\right)
\to \min(u,v)-uv.
\]
Notice that
$\max_{1\leq\ell\leq L}
{n_\ell^2/((n_\ell+2)}{V_L})\to 0,
$
so the multipole contributions are asymptotically negligible and the standard Lindeberg and tightness conditions follow from their Dirichlet fourth-moment bounds and the independent triangular array of multipole contributions.  Hence, $Z_L(\cdot)$ converges  the Brownian bridge $B(\cdot)$  in the Skorokhod space $D[0,1]$, see \cite{Billingsley1999}.
Then, one can use  the corresponding Kolmogorov-Smirnov statistic
$
T_L^{\mathrm{KS}}
=
\sup_{0\le u\le1}|Z_L(u)|.
$ Large values of $T_L^{\mathrm{KS}}$ therefore provide evidence against statistical isotropy within the diagonal model.
Under $H_0$, when $L\to \infty$, one can use the asymptotic result
\begin{equation*}
T_L^{\mathrm{KS}}
\Rightarrow
\sup_{0\le u\le1}|B(u)|
\end{equation*}
and the asymptotic tail probability, which is given by (\cite{Marsaglia2003})
\begin{equation*}
\mathbb P\left(\sup_{0\le u\le1}|B(u)|>x\right)
=
2\sum_{j=1}^{\infty}(-1)^{j-1}e^{-2j^2x^2},
\quad x>0.
\label{eq:kolmogorovtail-sphere}
\end{equation*}

\medskip

\noindent{\bf Consistency of the CUSUM test.}
Suppose that the alternative $H_1$ holds with a positive non-constant continuous function $g(\cdot)$ on $[0,1]$.

Let us denote
\[
S_\ell(u):=\frac1{n_\ell}\sum_{m=0}^{\lfloor\ell u\rfloor}
g(m/\ell)Z_{\ell m}.
\]
Continuity of the positive function $g(\cdot)$ on $[0,1]$ implies that it is bounded and also bounded away from zero. Since $\mathbb E(Z_{\ell m})=d_m$ and $\operatorname{Var}(Z_{\ell m})=2d_m$, when $\ell\to\infty,$
\begin{align*}
\mathbb E S_\ell(u)
&=\frac1{n_\ell}\sum_{m=0}^{\lfloor\ell u\rfloor}d_mg(m/\ell)
\longrightarrow\int_0^u g(x)\,dx,\\
\operatorname{Var}\left(S_\ell(u)\right)
&=\frac2{n_\ell^2}\sum_{m=0}^{\lfloor\ell u\rfloor}d_mg(m/\ell)^2
=O(n_\ell^{-1}).
\end{align*}
Therefore, $S_\ell(u)\overset p\to\int_0^u g(x)\,dx$ and
$S_\ell(1)\overset p\to\int_0^1g(x)\,dx>0$. Noting that
$F_\ell(u)=S_\ell(u)/S_\ell(1)$ and $p_\ell(u)\to u$, Slutsky's theorem gives
for each fixed $u\in[0,1]$,
\begin{equation}\label{eq:Glimit-sphere}
F_\ell(u) - p_\ell(u)
\overset{p}{\to}
G(u)-u
:=
\frac{\int_0^u g(x)\,dx}
     {\int_0^1 g(x)\,dx}-u.
\end{equation}

Let
\[
\overline F_L(u):=\frac1{L(L+2)}\sum_{\ell=1}^Ln_\ell F_\ell(u)
\qquad\mbox{and}\quad
\overline p_L(u):=\frac1{L(L+2)}\sum_{\ell=1}^Ln_\ell p_\ell(u).
\]
For each fixed $u\in[0,1]$, (\ref{eq:Glimit-sphere}), boundedness of $F_\ell(u)$ ($0\leq F_\ell(u)\leq1$), and Toeplitz's lemma imply
$\mathbb E\overline F_L(u)\to G(u)$. Independence for different multipoles gives
\[
\operatorname{Var}\left(\overline F_L(u)\right)
\leq\frac1{4(L(L+2))^2}\sum_{\ell=1}^Ln_\ell^2=O(L^{-1}),
\]
so $\overline F_L(u)\overset p\to G(u)$, when $L\to \infty.$

Since $\overline F_L(\cdot)$ is non-decreasing, $G(\cdot)$ is continuous, and
\[
\frac{Z_L(u)}{\sqrt{L(L+2)/2}}
=\frac{A_L(u)}{L(L+2)}\sqrt{\frac{L(L+2)}{V_L}},
\]
by using (\ref{pu}) and $V_L/(L(L+2))\to1$, one obtains
\begin{equation}
\frac{Z_L(u)}{\sqrt{L(L+2)/2}}
\overset{p}{\longrightarrow}
G(u)-u.
\label{eq:alternativegrowth-sphere}
\end{equation}
If the function $g(\cdot)$ is continuous and nonconstant, then $G(u)\neq u$ at some subinterval of $(0,1)$. Therefore, by \eqref{eq:alternativegrowth-sphere},
\begin{equation*}
\frac{T_L^{\mathrm{KS}}}{\sqrt{L(L+2)/2}}
\overset{p}{\longrightarrow}
\sup_{0\le u\le1}|G(u)-u|>0,
\label{eq:KS-consistency}
\end{equation*}
when $L\to \infty.$

Hence, $T_L^{\mathrm{KS}}\to\infty$ in probability. Thus, the CUSUM test is consistent against every fixed positive nonconstant alternative satisfying the stated regularity conditions.

\medskip

\noindent{\bf Finite-sample calibration and simulation.}
For small and moderate $L$, simulation from the exact null law can be preferable to using the Brownian-bridge approximation.  For each Monte Carlo replication $s=1,\ldots,S,$ and each $\ell=1,\ldots,L$, one can independently generate
$
(P^{(s)}_{\ell0},\ldots,P^{(s)}_{\ell\ell})
\sim
\operatorname{Dirichlet}(1/2,1,\ldots,1),
$
and then construct~$Z_L^{(s)}$ from~\eqref{eq:ZL}.

Denoting
\[
T_{L,s}
:=
\sup_{0\le u\le1}|Z_L^{(s)}(u)| \quad \mbox{and}\quad T_L^{\mathrm{obs}}=\sup_{0\leq u\leq1}|Z_L^{\mathrm{obs}}(u)|
\]
the Monte Carlo CUSUM $p$-value can be computed as
\begin{equation*}
\widehat p_{c,MC}
:=
\frac{1+
\sum_{s=1}^{S}
\mathds{1}\left(T_{L,s}\ge T_L^{\mathrm{obs}}\right)}
{S+1}.
\label{eq:mcP-sphere}
\end{equation*}

\section{Numerical studies}\label{p3sec5}

The numerical studies were conducted using Python~3.14.7, the \texttt{healpy} package~\citep{zonca2019}, and publicly available CMB data (see Appendix~\ref{secA1}).

For numerical studies, we used three types of data. First, we simulated harmonic coefficients $a_{\ell m}$ of an isotropic Gaussian spherical random field with an angular power spectrum
$
C_\ell = {1}/{(\ell(\ell+1))}, $ $ \ell > 0.
$
The field realisations were obtained on a HEALPix grid (see Appendix~\ref{secA2}) with the resolution parameter $N_{\mathrm{side}}=2048$, corresponding to
$N_{\mathrm{pix}}=50{,}331{,}648$ pixels. The associated HEALPix multipole limit is $\ell_{\max}=6143$. To ensure computational efficiency and numerical reliability, the score tests used multipoles up to $L=512$.
Second, we simulated an anisotropic random field by transforming the harmonic coefficients of the mentioned isotropic field as
$\tilde{a}_{\ell m}=a_{\ell m}\cdot\sqrt{0.1+{m}^2/{\ell}^2}.$ The additive term $0.1$ was included to prevent degenerate values of $\tilde{a}_{\ell m}$ when $m=0$.
The resulting field has independent harmonic coefficients $\tilde{a}_{\ell m}$ with the variances
$\operatorname{Var}(\tilde{a}_{\ell m})= C_\ell\cdot \left(0.1+{m}^2/{\ell}^2\right),$
which depend on the azimuthal order $m$. Finally, we used the Planck SMICA CMB data, see the details in Appendix~\ref{secA1}.

\begin{figure}[htbp]
    \centering
    \begin{subfigure}[b]{0.48\textwidth}
        \centering
        \includegraphics[width=\textwidth]{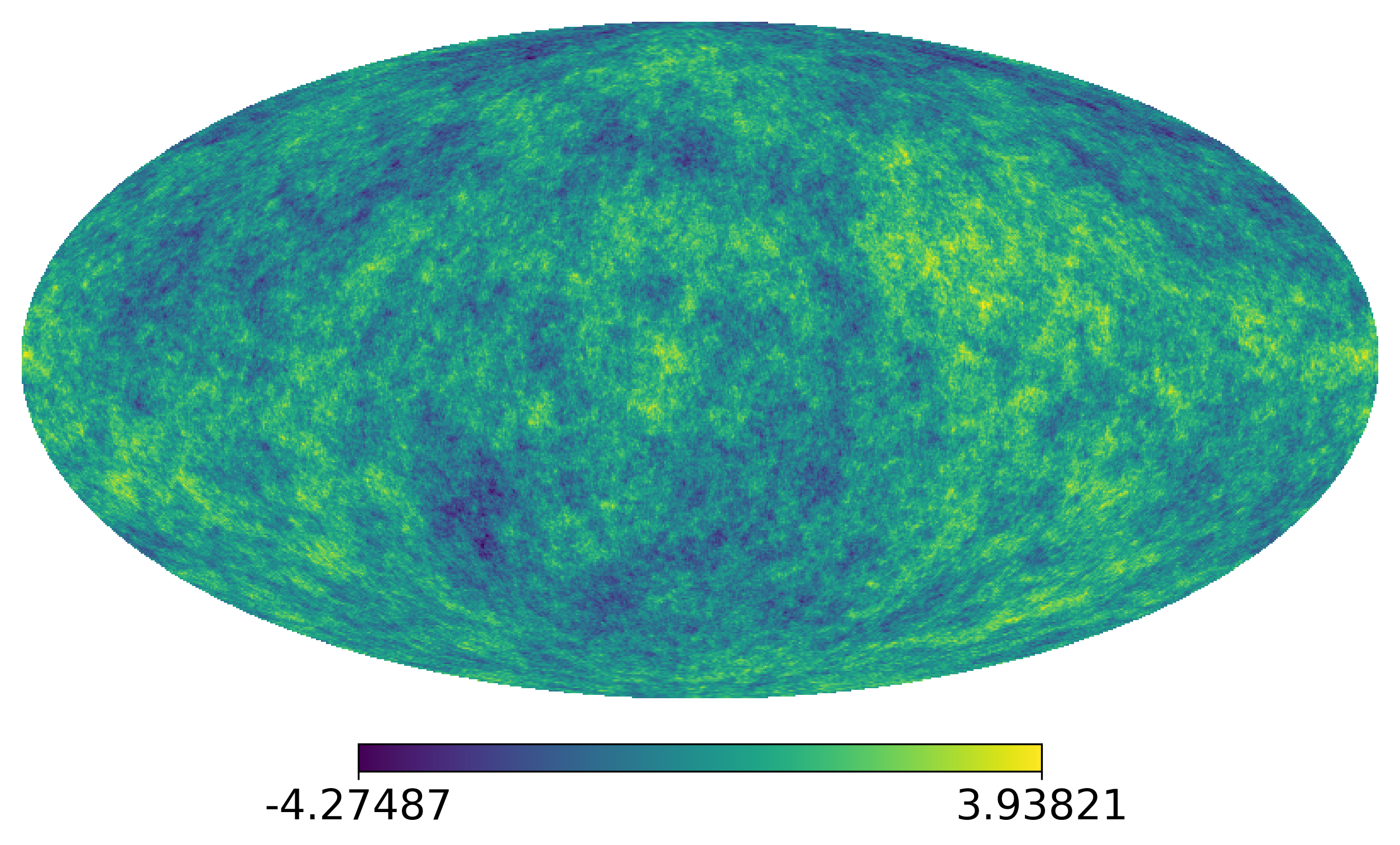}
        \caption{Isotropic field with $a_{\ell m}$ corresponding to $C_\ell={1}/(\ell(\ell+1))$ }
        \label{p3image2a}
    \end{subfigure}
    \hfill
    \begin{subfigure}[b]{0.48\textwidth}
        \centering
        \includegraphics[width=\textwidth]{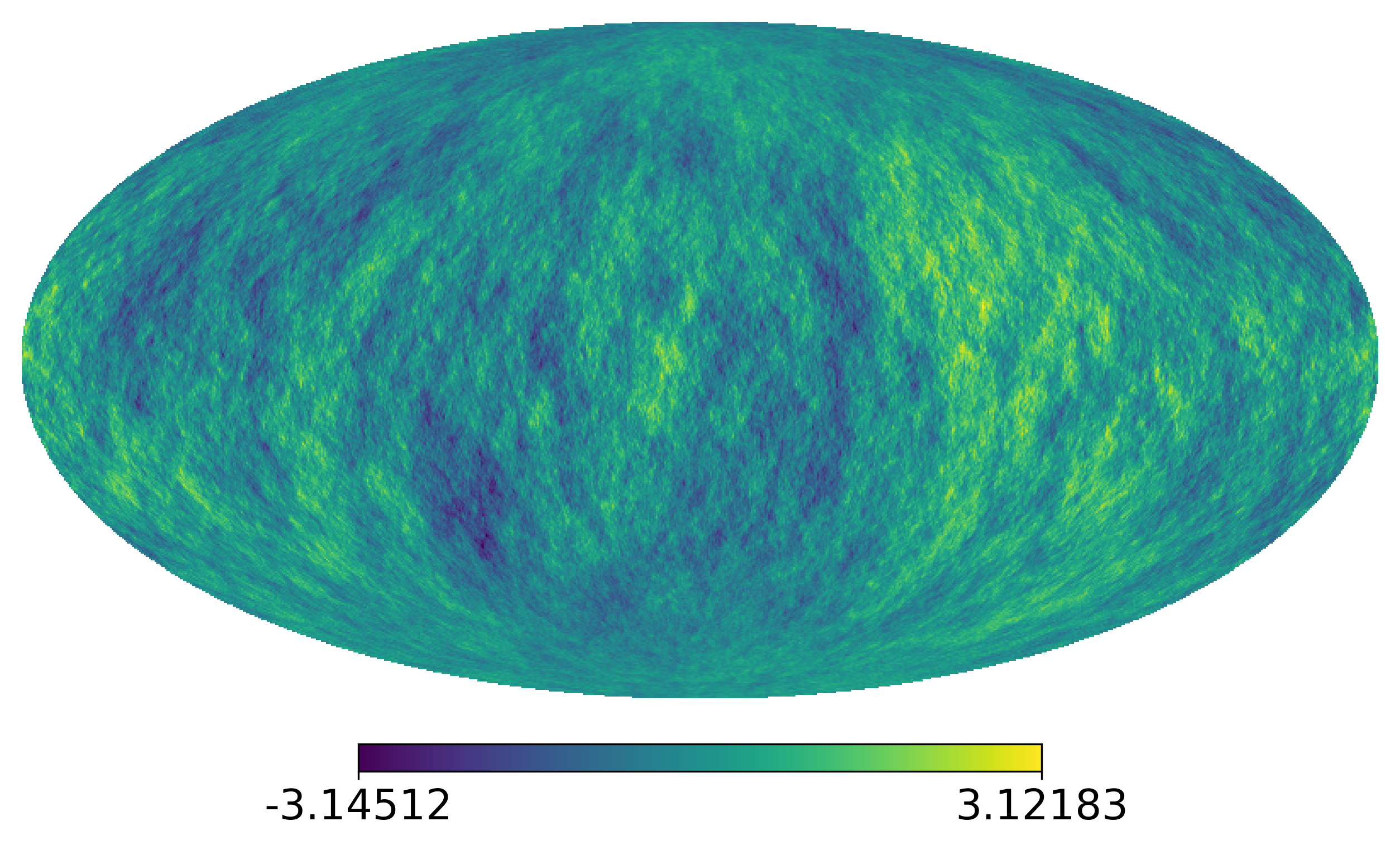}
        \caption{Anisotropic field with harmonic coefficients $\tilde{a}_{\ell m}=a_{\ell m}\cdot \sqrt{0.1+{m}^2/{\ell}^2}$}
        \label{p3image2b}
    \end{subfigure}

    \caption{Simulated spherical random fields in Mollweide projection.}
    \label{p3image2}
\end{figure}

The same resolution 2048 and multipole ranges were used for the simulated fields and the CMB data. Planck CMB map is plotted in Figure~\ref{p3image1}. Figures~\ref{p3image2a} and~\ref{p3image2b} show realisations of the mentioned isotropic and diagonal anisotropic spherical random fields, respectively. The fields are displayed using the Mollweide projection of the sphere, see \cite{snyder1987}.

For these numerical studies, three SMICA CMB intensity maps at resolution $N_{\mathrm{side}}=2048$ were used: the Planck 2015 release map (field 0, version R2.01), the Planck 2018 legacy release map (field 0, version R3.00), and the inpainted intensity map from the 2018 release (field 5, version R3.00). Each map was constructed by combining data from the nine Planck frequency channels of the low- and high-frequency instruments.

The Milky Way’s stars, gas, and dust produce strong galactic foreground emission that contaminates the CMB signal. The corresponding pixels are excluded by confidence masks, which remove up to $20\%$ of the sky, depending on the method and mask used. Outside the masked regions, the map values are retained. For the two field-0 maps, masked regions are filled during the SMICA processing using diffusive inpainting, followed by harmonic synthesis (\cite{planck2018}).

The inpainted field-5 intensity map is shown in Figure~\ref{p3image1}. In this map, the masked regions are replaced with a constrained Gaussian realisation drawn from the conditional distribution of the CMB field given the unmasked data. Thus, field-5 data are obtained via an additional simulation step, whereas the field-0 maps do not. The resulting field-5 values provide a statistically consistent realisation of the CMB in the masked regions under the assumed isotropic model \cite{planck2018}.

For the score test, we used $K=1,\ldots,10$ basis functions based on the shifted Legendre polynomials
$
h^{(k)}(x)=P_k(2x-1).
$
Monte Carlo estimates of the $p$-values were obtained from $S=1000$ simulations, using a significance level of $\alpha=0.05$.

\begin{figure}[htbp]
    \centering
    \begin{subfigure}[b]{0.48\textwidth}
        \centering
        \includegraphics[width=\textwidth]{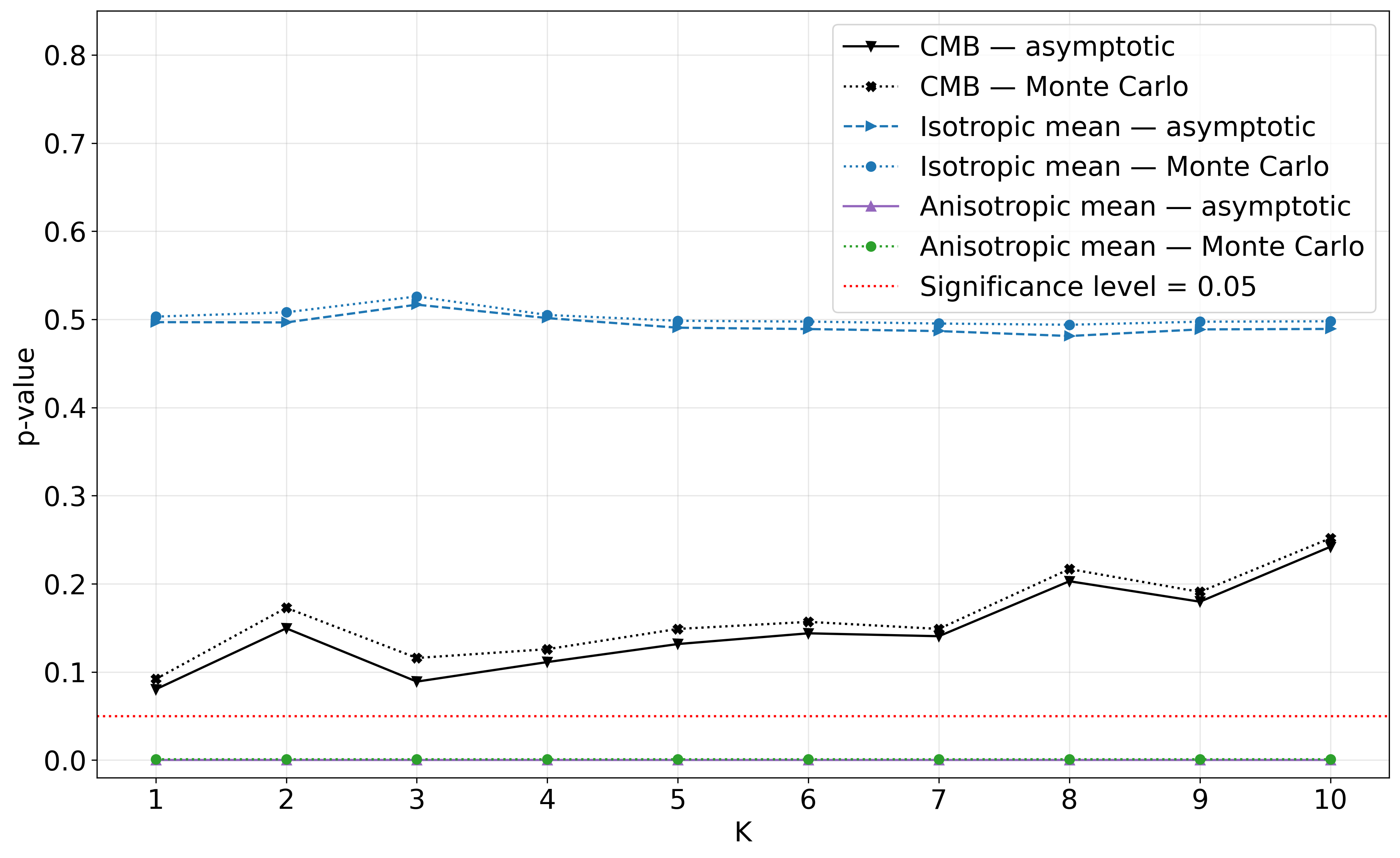}
        \caption{Score test}
        \label{p3image3a}
    \end{subfigure}
    \hfill
    \begin{subfigure}[b]{0.48\textwidth}
        \centering
        \includegraphics[width=\textwidth]{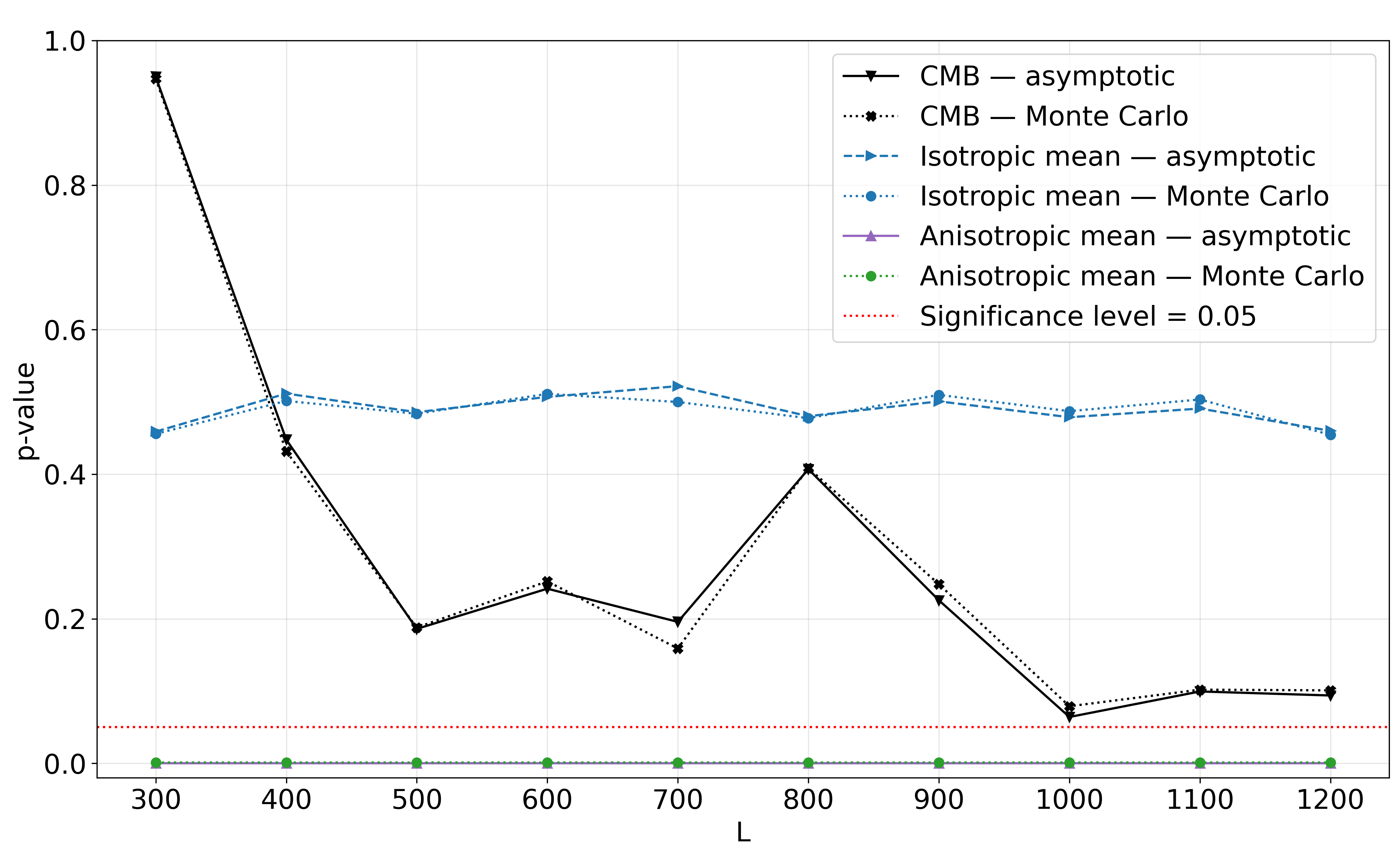}
        \caption{CUSUM test}
        \label{p3image3b}
    \end{subfigure}

    \caption{$p$-values for the simulated fields and CMB data (field 5, version R3.00).}
    \label{p3image3}
\end{figure}

Figure~\ref{p3image3a} presents the $p$-values of the score test as a function of the number of basis functions for the inpainted field-5 CMB data and the simulated isotropic and anisotropic fields at the same resolution. The asymptotic and Monte Carlo $p$-values are in close agreement, confirming the accuracy of the asymptotic approximation for moderate and large multipoles.

To obtain the top lines in Figure~\ref{p3image3a}, we generated 100 realisations of the isotropic random field, similar to the realisation shown in Figure~\ref{p3image2a}. For each $K$, the corresponding average $p$-values for these realisations were computed. It fluctuated between approximately $0.48$ and $0.52$, remaining well above the specified significance level. This is consistent with the expected behaviour of the score test for isotropic Gaussian fields.

Similar lines were produced for simulated anisotropic fields, such as the realisation shown in Figure~\ref{p3image2b}. The strong departure from isotropy resulted in the average $p$-values effectively equal to zero. The lower curve in Figure~\ref{p3image3a} shows that the score test consistently rejects the null hypothesis of isotropy for all values of $K$.

Using the score test at $L=512$, the asymptotic $p$-values for the inpainted data were between $0.080$ and $0.242$, in agreement with the Monte Carlo $p$-values. Thus, isotropy is not rejected for any considered set of basis functions. This result may be attributable to the masked pixels having been replaced by a constrained Gaussian realisation drawn from the conditional distribution under an isotropic CMB model.

Figure~\ref{p3image3b} shows the $p$-values of the CUSUM test as a function of the maximum multipole $L$ for the same field-5 CMB data and simulated isotropic and anisotropic fields. For the 100 simulated isotropic fields, the mean $p$-values remain between $0.46$ and $0.52$ for all values of $L$, so isotropy is not rejected. For the simulated anisotropic fields, the average $p$-values are close to zero, and isotropy is rejected in every case.

However, when the CUSUM test is applied to the inpainted field-5 CMB data, the $p$-values are not stable and vary noticeably with the value of~$L$. This variation may result from the smaller number of $a_{\ell m}$ coefficients available at low multipoles, as well as sampling variability associated with a single realisation of the observed CMB sky. Nevertheless, for all values of $L$, the CUSUM test provides no statistically significant evidence against isotropy.

Table~\ref{tab1P3} reports the rejection rates obtained from the 100 simulated fields using asymptotic and Monte Carlo $p$-values for the score test. For realisations simulated from the isotropic model, the rejection rates remain close to the nominal significance level, $\alpha=0.05$, ranging from $2\%$ to $7\%$. The CUSUM test yielded results with slightly higher percentages, see Table~\ref{ta2P3}. The small rejection $p$-values for $L=500$ motivated the selection of $L=512$.  For data simulated from the anisotropic model, both tests reject isotropy in $100\%$ of the simulations, for both computational methods.

\begin{table}[htb]
\centering
\captionsetup{width=\textwidth}
\caption{Empirical rejection rates (in \%) of the score test for isotropy}
\label{tab1P3}
\setlength{\tabcolsep}{8pt}
\begin{tabular}{@{}lcccccccccc@{}}
\toprule
$K$ & 1 & 2 & 3 & 4 & 5 & 6 & 7 & 8 & 9 & 10 \\
\midrule
Asymptotic  & 4 & 3 & 4 & 2 & 3 & 4 & 7 & 7 & 7 & 5 \\
Monte Carlo & 2 & 3 & 3 & 2 & 3 & 4 & 6 & 4 & 5 & 3 \\
\bottomrule
\end{tabular}
\end{table}

\begin{table}[htb]
\centering
\captionsetup{width=\textwidth}
\caption{Empirical rejection rates (in \%) of the CUSUM test for isotropy}
\label{ta2P3}
\setlength{\tabcolsep}{8pt}
\begin{tabular}{@{}lcccccccccc@{}}
\toprule
$L$ & 300 & 400 & 500 & 600 & 700 & 800 & 900 & 1000 & 1100 & 1200 \\
\midrule
Asymptotic  & 8 & 9 & 2 & 6 & 5 & 7 & 8 & 3 & 3 & 7 \\
Monte Carlo & 8 & 9 & 2 & 6 & 5 & 7 & 6 & 2 & 2 & 7 \\
\bottomrule
\end{tabular}
\end{table}


Notice, that failure to reject isotropy based on this CMB data does not constitute conclusive evidence that the observed CMB field is isotropic, because the masked values were imputed from the conditional distribution of an isotropic Gaussian model, and the analysis considered only a specific diagonal form of anisotropic alternative.

\begin{figure}[htbp]
    \centering
    \begin{subfigure}[b]{0.48\textwidth}
        \centering
        \includegraphics[width=\textwidth]{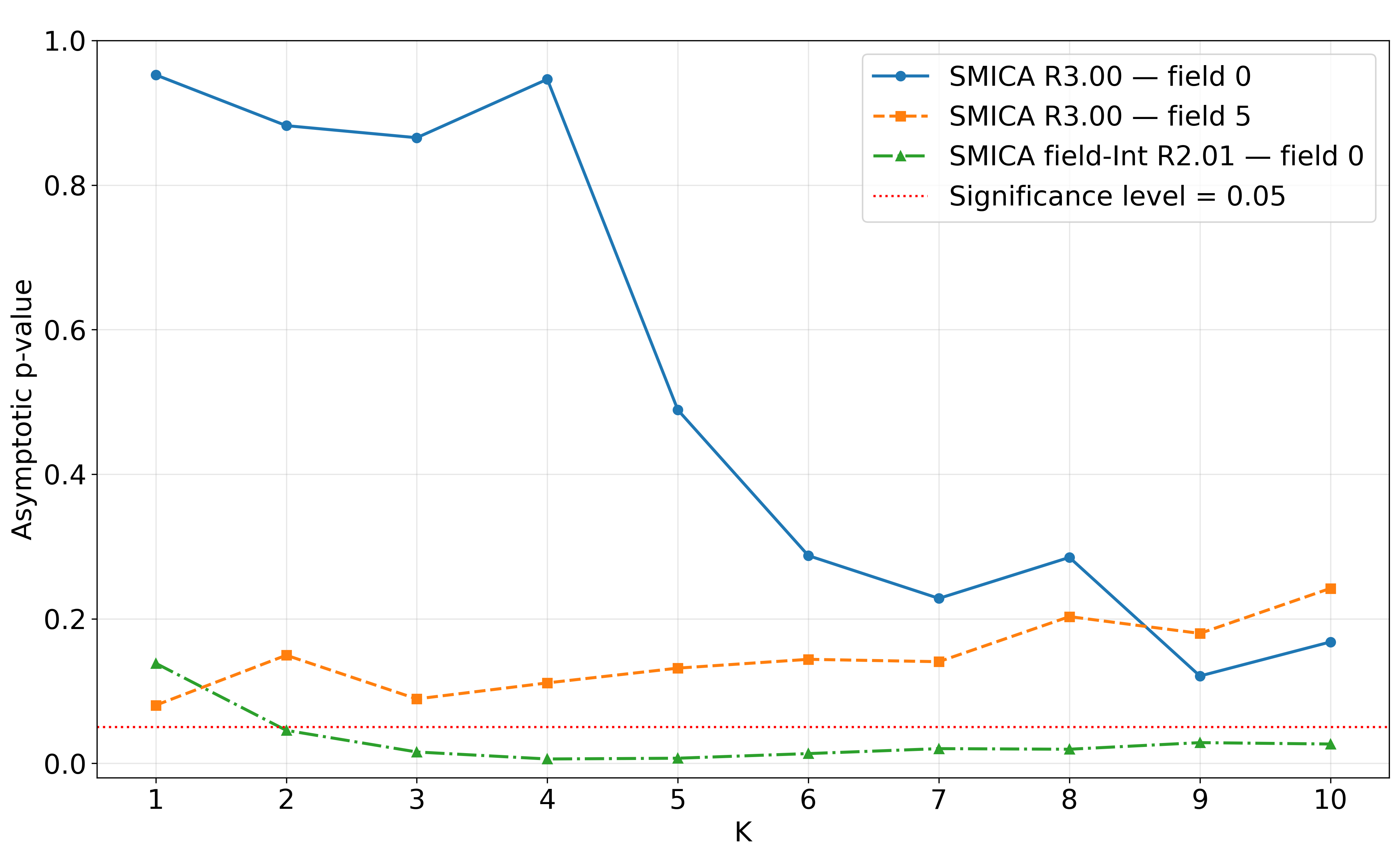}
        \caption{Score test}
        \label{p3image4a}
    \end{subfigure}
    \hfill
    \begin{subfigure}[b]{0.48\textwidth}
        \centering
        \includegraphics[width=\textwidth]{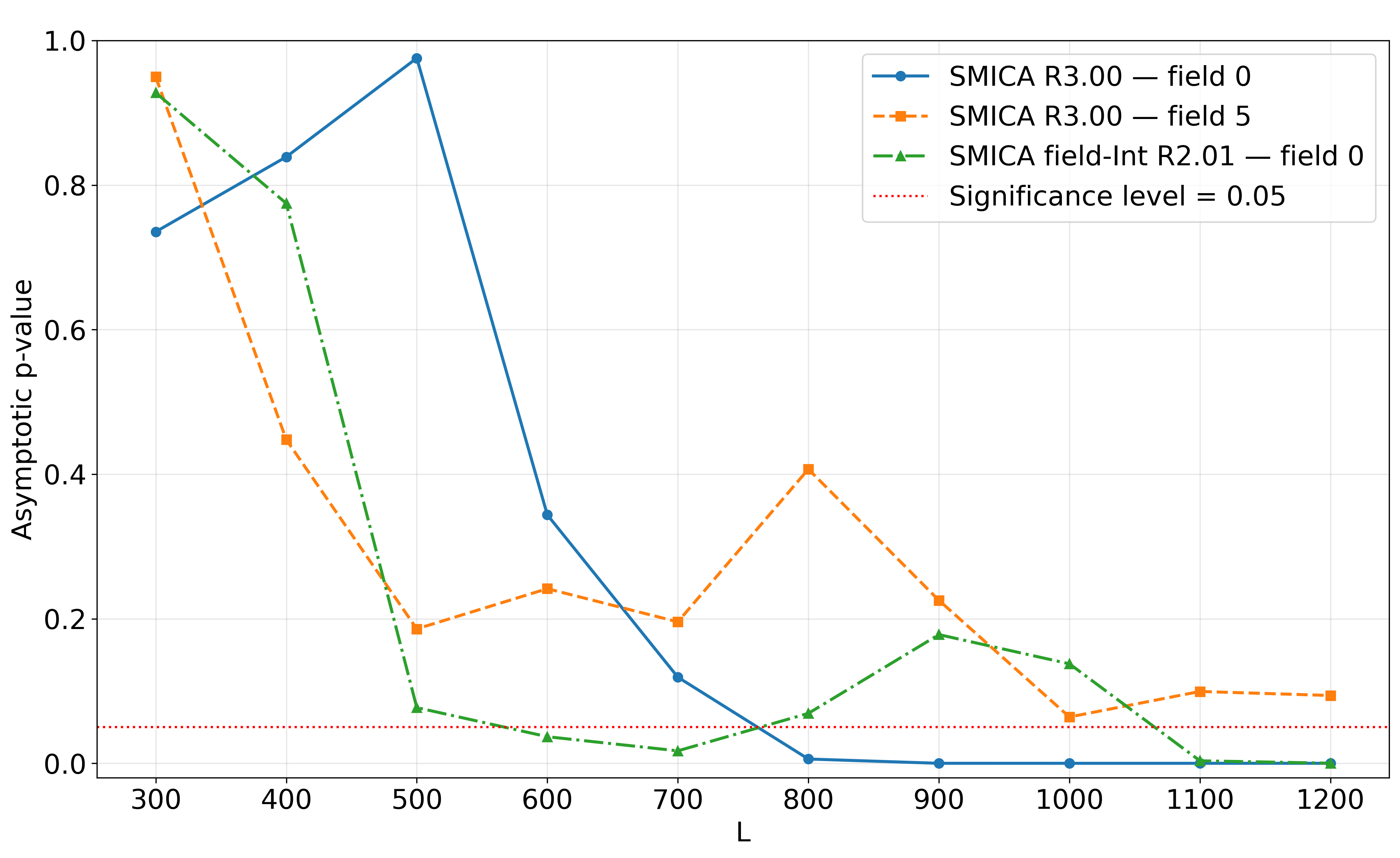}
        \caption{CUSUM test}
        \label{p3image4b}
    \end{subfigure}
    \caption{Asymptotic $p$-values for three SMICA CMB intensity maps.}
    \label{p3image4}
\end{figure}

We also compared the three considered CMB datasets to assess the performance of the tests for different maps. Figure~\ref{p3image4a} presents the $p$-values of the score test as a function of the number of basis functions for the three CMB maps. Neither of the two 2018 maps rejects isotropy at any value of $K$. The inpainted field-5 data remain stable as additional basis functions are included, with no abrupt fluctuations. In contrast, the 2015 map falls below the significance level at $K=2$, and isotropy is rejected for all subsequent values of $K$. This difference in the test results may reflect changes between the data releases in R2.01 and R3.00, especially their reconstruction in the masked regions.

Figure~\ref{p3image4b} compares the $p$-values of the CUSUM test as a function of the maximum multipole $L$ for the three Planck CMB maps considered. Again, the $p$-values for the inpainted field-5 2018 map remain above the significance level for all values of $L$. For the 2018 field-0 map, the $p$-values fall below the significance level from approximately $L=800$ onward, whereas the 2015 map $p$-values fall below the threshold at several values of $L$. These results may reflect differences between the field-0 and field-5 map construction procedures and between the R2.01 and R3.00 data releases. Substantial fluctuations in the $p$-values as a function of $L$ are observed for all three datasets and would require further investigation.

\section{Conclusion}\label{p3sec6}

This paper develops tests for isotropy against diagonal anisotropy, under which the harmonic coefficients remain uncorrelated, while their variances may vary with the azimuthal order. It introduces a semiparametric model in which these variances share a common modulation function. Then, under infill sampling, testing isotropy can be formulated as a test of homoscedasticity. Two procedures (a score test and a cumulative sum test) are proposed. Their asymptotic distributions and theoretical properties are derived when the number of multipole degrees increases. We also provided exact-null Monte Carlo calibration for moderate multipole ranges. The developed tests were applied to simulated fields and actual Planck CMB data.

Several extensions would be interesting for further investigation. One direction is to develop statistical methods for other classes of anisotropic models introduced in~\cite{Durastanti2026}. Future work could also evaluate the tests for a broader range of settings and establish appropriate procedures for selecting their parameters. It would be important to further develop the methodology and its applications to real spherical datasets arising in cosmology and the geosciences.

\backmatter

\bmhead{Supplementary information}

The corresponding Python code is freely available in the folder
”Research materials” from the website \url{https://sites.google.com/site/olenkoandriy/}.

\bmhead{Acknowledgements}
 A. Olenko would like to express his gratitude for the hospitality provided by Prof.~C.Durastanti and Sapienza Universit\`a di Roma during his sabbatical in~2024, that contributed to the initiation of this research.

During the preparation of this manuscript, the authors used GPT-5.6 Sol to optimise Python code and improve the language. All final code and text were reviewed and verified by the authors.

\section*{Declarations}

\begin{itemize}
\item Funding

This research was supported under the Australian Research Council's Discovery Projects funding scheme (project number  DP220101680). A.Olenko was partially supported by La Trobe University's SCEMS CaRE and Beyond grant.

\item Data availability

Freely available Planck SMICA CMB map \url{https://irsa.ipac.caltech.edu/data/Planck/release_2/all-sky-maps/matrix_cmb.html} and \url{https://irsa.ipac.caltech.edu/data/Planck/release_3/all-sky-maps/matrix_cmb.html}.
\end{itemize}

\bigskip

\begin{appendices}

\section{CMB background and data}\label{secA1}
The Universe began approximately 13.8 billion years ago in a very hot and dense state. As the Universe started expanding, it slowly cooled off, and around 380,000 years after the Big Bang, atoms were formed. This recombination allowed photons to travel freely through space. This radiation is observed today as the CMB. The CMB is believed to be the oldest observable radiation and a leftover from the early Universe. This radiation provides strong support in favour of the Big Bang theory. The CMB is almost uniform over the sky with a temperature of approximately $2.73\,\mathrm{K}$, but shows small variations in its temperature, which provide important information for modern cosmological theories. Traditionally, the CMB is modelled as a spherical random field and analysed using spherical harmonics~\cite{marinucci2011}. In this work we use the freely available Planck SMICA CMB data \url{https://irsa.ipac.caltech.edu/data/Planck/release_2/all-sky-maps/matrix_cmb.html} and \url{https://irsa.ipac.caltech.edu/data/Planck/release_3/all-sky-maps/matrix_cmb.html}, which provide CMB measurements and their reconstructions in the masted regions.

\section{HEALPix pixelisation of sphere}\label{secA2}

The HEALPix pixelisation is a uniform scheme which provides an efficient way to organise and represent spherical data specially the CMB radiations. The CMB observation in HEALPix format is stored as Flexible Image Transport System (FITS) files, where each pixel has a unique location that may contain temperature intensity and polarisation information. HEALPix divides the sphere into equal-area pixels.

The HEALPix scheme start with by dividing the unit sphere into 12 equal-area quadrilateral base pixels arranged over three regions. For the initial resolution parameter ($j=0$), these form the base pixel structure. For each \(j>0\), every pixel is subdivided into four equal-area quadrilateral child pixels until the required resolution is achieved. The total number of pixels on the sphere is obtained from this grid resolution is $N_{\text{pix}}= 12N_{\text{side}}^2$, where
$N_{\text{side}} = 2^j,$ $ j \in \mathbb{N}_0,$ is the grid resolution. The upper limit for the spherical harmonic can be computed by $\ell_{\max} = 3N_{\text{side}} - 1,$ which gives the highest multipole supported by the HEALPix resolution \cite{gorski2005}.
Therefore, increasing $j$, and consequently $N_{\mathrm{side}}$, increases the number of equal-area pixels and provides a finer representation of data on the sphere.




\end{appendices}


\bibliography{biblography}

\end{document}